\documentclass[graybox]{svmult}

\usepackage{mathptmx}       
\usepackage{helvet}         
\usepackage{courier}        
\usepackage{type1cm}        
\usepackage{makeidx}         
\usepackage{graphicx}        
\usepackage{multicol}        
\usepackage[bottom]{footmisc}

\usepackage[authoryear,longnamesfirst]{natbib}
\bibpunct{(}{)}{;}{a}{}{,}

\makeindex             

\begin{document}

\title*{Supernova of 1006 (G327.6+14.6)}
\titlerunning{Supernova of 1006 (G327.6+14.6)}
\author{Satoru Katsuda}
\institute{Department of Physics, Faculty of Science \& Engineering, Chuo University, 1-13-27 Kasuga, Bunkyo, Tokyo 112-8551, Japan, \email{katsuda@phys.chuo-u.ac.jp}
}
%
%
\maketitle

\abstract{SN~1006 (G327.6+14.6) was the brightest supernova (SN) witnessed in human history.  As of one thousand years later, it stands out as an ideal laboratory to study Type Ia SNe and shocks in supernova remnants (SNRs).  The present state of knowledge about SN~1006 is reviewed in this article.  No star consistent with a surviving companion expected in the traditional single-degenerate scenario has been found, which favors a double-degenerate scenario for the progenitor of SN~1006.  Both unshocked and shocked SN ejecta have been probed through absorption lines in ultraviolet spectra of background sources and thermal X-ray emission, respectively.  The absorption studies suggest that the amount of iron is $<$0.16\,M$_\odot$, which is significantly less than the range for normal SNe Ia.  On the other hand, analyses of X-ray data reveal the distribution of shocked ejecta to be highly asymmetric especially for iron.  Therefore, most of iron might have escaped from the ultraviolet background sources.  Another important aspect with SN~1006 is that it was the first SNR in which synchrotron X-ray emission was detected from shells of the remnant, providing evidence that electrons are accelerated up to $\sim$100\,TeV energies at forward shocks.  The bilateral symmetry of the synchrotron emission (bright in northeastern and southwestern limbs) is likely due to a polar cap geometry.  The broadband (radio, X-ray, and gamma-ray) spectral energy distribution suggests that the gamma-ray emission is predominantly leptonic.  At the northwestern shock, evidence for extreme, but less than mass proportional, temperature non-equilibration has been found by optical, ultraviolet, and X-ray observations.}

\section{Introduction}
\label{sec:1}

SN~1006 (G327.6+14.6) was the brightest supernova (SN) recorded in human history.  According to various historical records ($\sim$30 in total from China, Japan, Korea, Europe, and the Arab world) and their different interpretations, the peak magnitude of SN~1006 ranges from $-2$ to $-10$ (e.g., \citet{Stephenson2010}).  Of these, the most straightforward estimate would be $-7.3$ to $-7.6$ \citep{Winkler2003}, based on a simple interpretation of the Egyptian astrologer Ali bin Ridwan's description: ``its size 2$\frac{1}{2}$ to 3 times the magnitude of Venus" and ``a little more than a quarter of the brightness of the Moon".  On the other hand, Stephenson (2010) paid more attention to information from China: ``It was so brilliant that one could scrutinize things (presumably nearby objects)."  Since one can start to discern objects only when the Moon is about five days old (approximate magnitude $-8.5$ or $-9$) in a clear and very dark sky, he concluded that the peak brightness was around $-8.5$.  Perhaps, a fair estimate would be somewhere between these two plausible evaluations.

\begin{figure}[h]
\begin{center}
\includegraphics[scale=2]{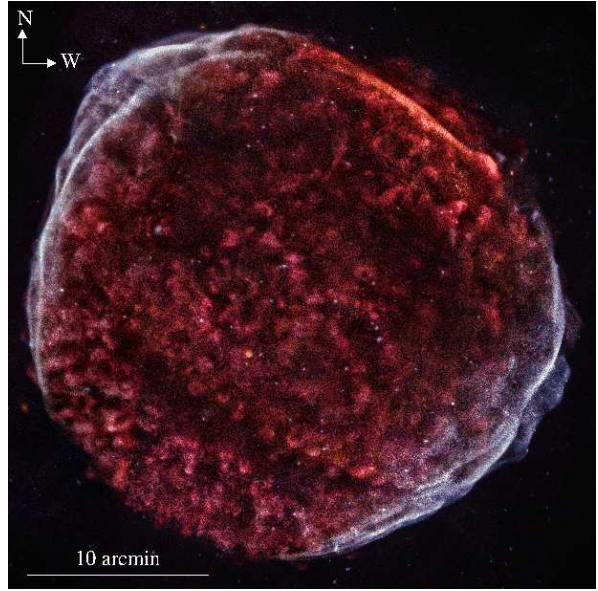}
\caption{Three-color X-ray image of the entire SN~1006 taken by {\it Chandra}.  Red, green, and blue are responsible for soft, medium, and hard X-rays.  This image was taken from the {\it Chandra} X-ray center.  }
\label{fig:1}       
\end{center}
\end{figure}

Today, it can be observed in many wavelengths as a shell-type supernova remnant (SNR) with a radius of 30$^{\prime}$, as shown in Fig.~\ref{fig:1} --- a deep X-ray image taken by {\it Chandra} in 2012 \citep{Winkler2014}.  SN~1006 was almost certainly a Type Ia event, based on its location far above the Galactic plane ($b=14.6^\circ$ = several hundreds of pc above the Galactic plane at a distance of 1--2\,kpc), the lack of both a nearby OB association and a compact remnant, and recorded visibility for nearly two years.  By combining the apparent magnitude of $-8$ with the absolute visual magnitude of normal SN Ia \citep[$-19.12$:][]{Leloudas2015} and the visual extinction of $A_{\rm v} = 0.31$ \citep{Winkler2003}, the distance is estimated to be only 1.45\,kpc.  It stands out among all historical SNRs as the closest, highest above the Galactic plane and thus least obscured, and largest.  It has been an excellent laboratory to study explosion mechanisms and nucleosynthesis of SNe Ia, as well as collisionless shock physics including cosmic ray acceleration.

\section{Searching for a Surviving Companion Star to Constrain a Progenitor System}
\label{sec:2}

There is a consensus that SNe Ia originate from thermonuclear explosions of C+O white dwarfs whose masses are close to the Chandrasekhar mass (1.38\,M$_\odot$).  However, it is not clear how normal C+O white dwarfs, born with $\sim$0.6\,M$_\odot$, approach the Chandrasekhar mass.  There are two main scenarios to increase the white dwarf mass: 1) the merger of two white dwarfs, i.e., the double-degenerate (DD) scenario \citep{Iben1984}, 2) mass accretion from a non-degenerate companion, i.e., the single-degenerate (SD) scenario \citep{Nomoto1982}.  The two scenarios predict a testable observational difference; nothing is left behind the explosion in the first channel, while a companion can survive the explosion in the second channel. 

Using Cerro Tololo Inter-American Observatory (CTIO) 4-m telescope, Schweizer \& Middleditch (1980) searched for a surviving companion star, and found an unusually hot subdwarf sdOB star (hereafter, the SM star) at 2.5$^\prime$ south from the geometric center of the SNR.  Since the distance to this object, 1.1$^{+1.4}_{-0.6}$\,kpc, was similar to SN~1006, they suggested that it may be the surviving companion star.  However, Savedoff \& van Horn (1982) argued against this interpretation, since the time for a surviving companion to cool to the observed temperature was much longer than the SNR age.  Also, the offset of 2.5$^\prime$ from the SNR center indicates a proper motion of $\sim$0.15$^{\prime\prime}$\,yr$^{-1}$, which was not detected.  Subsequently, using the {\it International Ultraviolet Explore} satellite, Wu et al. (1983) discovered broad absorption features corresponding to Fe II and Si II, III, IV in the UV spectrum of the SM star.  The large broadening clearly shows that they are not caused by the interstellar medium (ISM) but are due to freely expanding cold SN ejecta.  The fact that all of these absorption features have redshifted components means that the SM star should be located behind SN~1006, ruling out its physical associated with SN~1006.  Furthermore, Winkler et al. (2005) found two other background UV sources within SN~1006 that show similar broad absorption lines, which means that the SM star is not unique, and thus is undoubtedly not a stellar remnant of SN~1006.

Continuous searches for a stellar remnant have failed to detect viable candidates.  Gonz\'alez-Hern\'andez et al.\ (2012) found that all subgiants and main-sequence stars down to an absolute magnitude of the Sun ($L_\odot$) within a 4$^\prime$ radius of the apparent explosion site do not show any hints of surviving companions, i.e., noteworthy elemental abundances on the surface, radial velocity, and rotation.  Kerzendorf et al.\ (2012) performed a deeper optical survey within a 2$^\prime$ radius, scrutinizing all stars to a limit of 0.5\,$L_\odot(V)$ and performed radial velocity measurements of stars down to a limit of $\sim$0.1\,$L_\odot(V)$ at the distance of SN~1006.  The brightness limits investigated are well below theoretical expectations of surviving companion stars; giant donors or subgiant/main sequence donors will become $\sim$1000\,$L_\odot$ for at least 10$^5$\,yr (giant donor) or $\sim$500\,$L_\odot$ for 1400--11,000\,yr, respectively \citep{Marietta2000}.  Despite the complete samples, no candidate companion was found.  This is in conflict with the SD scenario, and favors the DD scenario or something else such as a spin-up/spin-down SD scenario in which the donor star evolved to a white dwarf before the SN explosion \citep{Hachisu2012}.

\section{Spectroscopic Studies of Unshocked and Shocked Ejecta}
\label{sec:3}

While detections of the broad UV absorption lines from the SM star revealed that it is not physically related to SN~1006 \citep{Wu1983}, these absorption features provide us with an unusual opportunity to probe unshocked ejecta within SNRs.  Therefore, extensive observational and theoretical studies have been performed for the SM star.  

\begin{figure}[h]
\begin{center}
\includegraphics[scale=3]{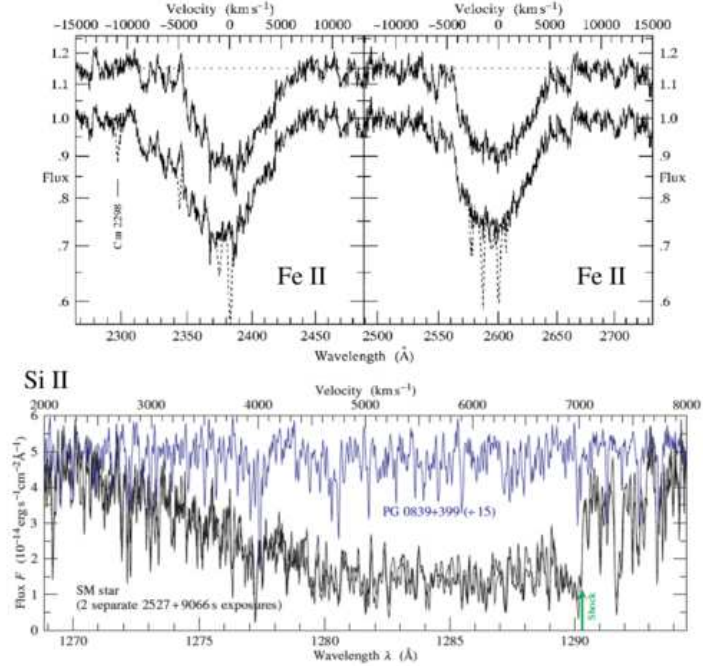}
\caption{Upper left panel: Fe II 2382.765, 2344.214, 2374.4612\,\AA\ absorption features in the SM star's spectrum.  The lower curve shows the dereddened spectrum, scaled to a continuum, with narrow interstellar Fe II, stellar C III 2297.578\,\AA\ lines subtracted as indicated by dashed lines.  The spectrum is an inverse-variance-weighted merger of {\it Hubble Space Telescope} G190H and G270H spectra.  Upper right panel: Same as left but for Fe II 2600.1729, 2586.6500\,\AA.  The narrow interstellar components are Mn II 2576.877, 2594.499, 2606.462\,\AA.  These are taken from Hamilton et al.\ (1997) with AAS permission.  Lower panel: STIS E140M spectra of the redshifted Si II 1260\,\AA\ feature of the SM star and the comparison star PG 0839+399. The spectrum of PG 0839+399 has been shifted by +32\,km\,s$^{-1}$ to mesh its stellar lines with those of the SM star, and divided by 15 to bring its continuum to approximately same level as the SM star.  All spectra have been smoothed with a near-Gaussian of FWHM 6.8\,km\,s$^{-1}$.  The vertical arrow shows the position of the putative reverse shock at 7026$\pm$10\,km\,s$^{-1}$.  This is taken from Hamilton et al.\ (2007).}
\label{fig:2}       
\end{center}
\end{figure}

As shown in Fig.~\ref{fig:2} upper panels \citep{Hamilton1997}, Fe II absorption lines at 2383\,\AA\ and 2600\,\AA\ are very broad ($\pm$5000\,km\,s$^{-1}$), and are centrally symmetric with respect to rest frames, suggesting that they are caused by symmetrically-distributed cold Fe inside the remnant.  The Fe II absorption profiles enable us to directly estimate a density profile of the singly-ionized Fe along the line of sight.  By integrating the density profile, Hamilton et al.\ (2007) calculated the mass of Fe$^+$ to be 0.029$\pm$0.004\,M$_\odot$, assuming spherical symmetric distribution of cold Fe.  Based on the fact that the ionization state of unshocked Fe appears to be quite low \citep{Blair1996}, with Fe$^+$/Fe being inferred to be 0.66$^{+0.29}_{-0.22}$, the total Fe mass was estimated to be 0.044$^{+0.022}_{-0.013}$\,M$_\odot$ with a 3$\sigma$ upper limit of 0.16\,M$_\odot$.  This is inconsistent with the expected presence of several tenths of a solar mass of Fe for normal SNe Ia.  

Different from the symmetric profiles of Fe II lines, Si absorption lines are entirely at redshifted velocities.  As an example, we show the Si II 1260\,\AA\ feature in Fig.~\ref{fig:2} lower panel \citep{Hamilton2007}.  The absence of the blueshifted Si absorption indicates no cold Si in the near side, suggesting a strong asymmetry in the ambient density along the line of sight in the sense that the far side is particularly tenuous compared with other regions of the remnant \citep{Hamilton1997}.  In addition, the sharp red edge at 7070\,km\,s$^{-1}$ indicates the position of the reverse shock.  This edge was later found to be shifting at a rate of $-4.2\pm$1.0\,km\,yr$^{-1}$, suggesting that the reverse shock is entering slower ejecta \citep{Winkler2011}.  Hamilton et al.\ (1997) interpreted that the Si II line consists of both unshocked and shocked Si II, whereas absorption features of Si III and Si IV do not show such a sharp edge, indicating that they arise mostly from shocked Si.  

\begin{figure}[h]
\begin{center}
\includegraphics[scale=.45]{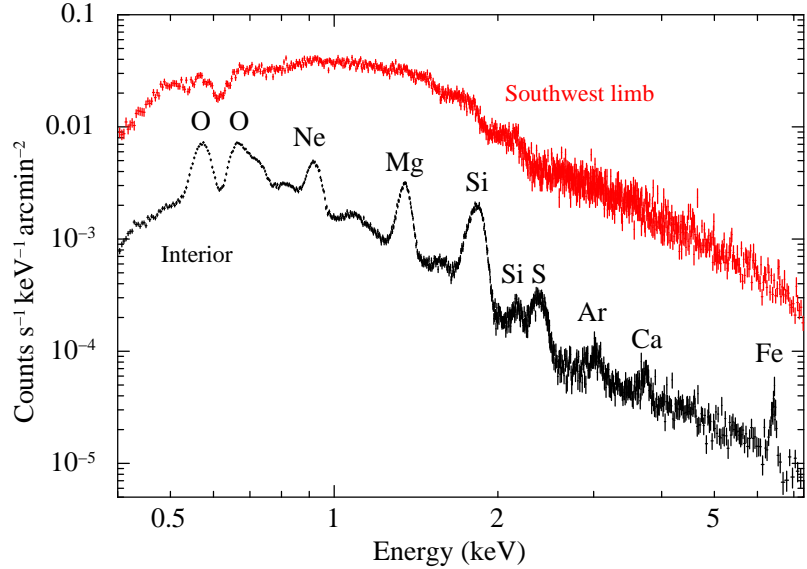}
\caption{X-ray spectra from the thermal (in black) and nonthermal (in red) dominated regions in SN~1006, taken by {\it Suzaku}.  Prominent lines are labelled by element.}
\label{fig:3}
\end{center}
\end{figure}

Apart from the UV spectroscopy used to probe the cold ejecta, X-ray spectroscopy is a strong tool to probe shocked ejecta, if the X-ray spectrum contains prominent line emission.  Becker et al.\ (1980) found that the X-ray spectrum from SN~1006 is featureless, which is in sharp contrast to line-dominated spectra seen in other shell-like SNRs such as Cassiopeia~A, Tycho, and Kepler.  This fact led the authors to suggest that the X-ray emission is predominantly nonthermal.  However, later X-ray observations started to show various indications of thermal emission.  These include O K-shell lines with the imaging gas scintillation proportional counter \citep{Vartanian1985}, a relatively steep slope of the hard X-ray spectrum compared with typical synchrotron nebula with {\it Tenma} \citep{Koyama1987}, and an Fe K-shell lines complex with the {\it Ginga} \citep{Ozaki1994} and the {\it BeppoSAX} satellites \citep{Vink2000}.  After these studies, the {\it ASCA} satellite finally settled the question of the nature of the X-ray emission \citep{Koyama1995}; bright northeastern (NE) and southwestern (SW) limbs are dominated by synchrotron radiation, while the faint interior regions are dominated by thermal emission.  Remarkable spectral differences between these two regions can be readily seen in Fig.~\ref{fig:3} --- X-ray spectra from the SW limb (in red) and the interior area (in black) acquired by the {\it Suzaku} satellite.

The spectrum from the interior region (black in Fig.~\ref{fig:3}) shows a number of lines from different species as labelled in the figure.  Their line center energies as well as the line intensity ratios indicate an extreme non-equilibrium ionization state \citep{Yamaguchi2008}, as is expected from the low environmental densities in the thermal-dominated limbs \citep{Acero2007,Winkler2014} as well as nonthermal-dominated limbs \citep{Korreck2004,Katsuda2009}.  Clumpy structures inside or outside the forward shock of the thermal emission suggests that it is dominated by the reverse-shocked SN ejecta rather than the swept-up ISM \citep{Cassam-chenai2008}, although some fractions (especially for O and Ne) must be attributed to the swept-up ISM \citep{Winkler2014}.  Yamaguchi et al.\ (2008) and Uchida et al.\ (2013) measured the relative abundances of shocked ejecta to be consistent with theoretical expectations based on standard SN Ia models.  They also found that the distribution of heavier elements, i.e., Si through Fe, is highly asymmetric, biased to the SE quadrant of the remnant, which was confirmed by {\it Chandra} and {\it XMM-Newton}'s deep followup observations \citep{Winkler2014,Li2015}.  The asymmetry would be caused by either the SN explosion asymmetry or the inhomogeneous ambient density.  If the former is the case, the deficiency of Fe inferred from UV absorption studies could be resolved; most of Fe would have been ejected toward the SE quadrant where the UV background sources are not present.  

\section{Cosmic Ray Acceleration at the Forward Shock}
\label{sec:4}

In the early 1950s, detections of nonthermal radio emission from SNRs established that SNRs contain nonthermal, power-law populations of relativistic electrons/positrons with energies of the GeV range.  About 40 years later, the {\it ASCA} and {\it ROSAT} satellites discovered the first nonthermal X-ray emission from the NE and SW limbs of SN~1006 \citep{Koyama1995,Willingale1996}.  This was the first evidence that SNR shocks can accelerate electrons up to TeV energies.  Since then, SN~1006 has been one of the most important SNRs to study cosmic ray acceleration such as mechanisms of diffusive shock acceleration (a widely accepted acceleration process at SNR shocks) and magnetic field amplification.

\subsection{Geometry of Synchrotron Emission}
\label{subsec:1}

The synchrotron surface brightness of SN~1006 shows a clear bilateral symmetry that the NE and SW limbs are particularly bright compared with other regions (cf.\ Fig.~\ref{fig:1}).  This indicates that the synchrotron geometry forms either polar caps or an equatorial belt.  In the polar cap model, the ambient magnetic field should be aligned NE--SW direction (parallel to the Galactic plane), while in the equatorial belt model it is aligned the SE--NW direction.  Since the shock obliquity, i.e., the angle between the shock normal and the magnetic field vector, plays a fundamental role in many aspects of SNR shocks, considerable efforts have been devoted to reveal the synchrotron geometry and the direction of the ambient magnetic field. 

It appears that the equatorial belt model was initially considered to be viable than the polar cap model \citep{Fulbright1990,Reynolds1996}, suggesting a large scale magnetic field aligned SE to NW.  However, one problem with this model is that the injection of thermal particles into acceleration is lower in the synchrotron-bright limbs than in the synchrotron-faint limbs, which conflicts to the observations.  The injection problem led V\"olk et al.\ (2003) to propose that the polar cap model is more reasonable than the equatorial belt model.  By performing spatially-resolved spectral analyses of the entire periphery of SN~1006 with {\it XMM-Newton} data, Rothenflug et al.\ (2004) found clear azimuthal variations of the cutoff frequency, ranging from 0.1\,keV at the synchrotron-bright limbs to 5\,keV at synchrotron-faint limbs.  The azimuthal variation was confirmed in more detail by followup X-ray observations \citep{Miceli2009,Katsuda2010}.  Such a strong variation can be explained only by the polar cap model, ruling out the equatorial belt model.  Another strong azimuthal variation was found in the ratio of radii between the forward shock and the contact discontinuity, based on the X-ray and optical data analyses \citep{Cassam-chenai2008}.  This is also in favor of the polar cap model.  Finally, Bocchino et al.\ (2011) compared an observed radio morphology at 1\,GHz with synthesized radio maps based on 3D magnetohydrodynamic (MHD) simulations, finding that a model assuming quasi-parallel injection efficiency better represents the data than the quasi-perpendicular model.  

\begin{figure}[h]
\begin{center}
\includegraphics[scale=.4]{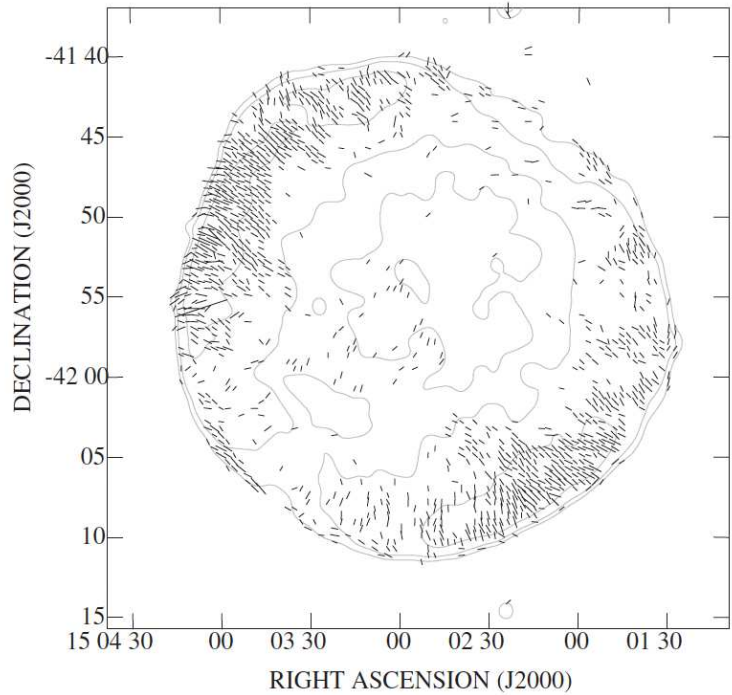}
\caption{Distribution of magnetic field vectors on SN~1006 at 1.4\,GHz, corrected for Faraday rotation (assuming uniform rotation measure = 12\,rad\,m$^{-2}$), at 10$^{\prime\prime}$ resolution.  For the vectors, a length of 30$^{\prime\prime}$ represents 0.26\,mJy\,beam$^{-1}$ of polarized flux.  Total intensity contours are superposed.  This is taken from Reynoso et al.\ (2013)  with AAS permission.}
\label{fig:4}
\end{center}
\end{figure}

Direct estimates of the magnetic-field orientation were performed through radio polarization measurements for the bright lobes \citep{Kundu1970}.  These observations consistently measured fractional polarization of 10--20\% as well as the orientation of the magnetic field aligned NE to SW.  Since this orientation is radial as is generally observed in young SNRs \citep{Dickel1976}, the result was not conclusive in determining the direction of the magnetic field around SN~1006.  Reynoso et al.\ (2013) performed sensitive polarimetric observations of the faint SE and NW limbs for the first time, finding highly polarized (60$\pm$20\%) radio emission as well as magnetic fields aligned tangent to the shock fronts as shown in Fig.~\ref{fig:4}.  This result strongly suggests that the ambient field is aligned this (SE--NW) direction, and that the bilateral synchrotron morphology is due to the polar cap geometry.  Recent 3D MHD simulations of polarized emission in SN~1006 also supported the quasi-parallel model \citep{Schneiter2015}.

\subsection{Magnetic Field Amplification and Efficient Cosmic Ray Acceleration}
\label{subsec:2}

It has been believed that SNRs are accelerating cosmic rays up to the ``knee", the slight inflection and steepening around $3\times10^{15}$\,eV.  In the frame of the diffusive shock acceleration theory, strong magnetic fields compared with the interstellar value of a few $\mu$G are required for rapid acceleration that allows particles to reach the knee energy.  

While magnetic field strengths in SNRs have never been directly measured so far, recent X-ray observations enabled us to indirectly estimate the field strength.  For instance, the {\it Chandra} X-ray observatory has revealed remarkably narrow synchrotron rims at SNR shells \citep{Bamba2003}.  To explain the narrowness, one must either eliminate the radiating particles or drop the magnetic field downstream.  The former scenario had been considered from the early studies, in which an electron could only travel a certain distance before loosing its energy due to synchrotron losses.  This distance is determined by two competing transport mechanisms: advection (the flow of the plasma) or diffusion (random motion of electrons on the scale of gyroradii).  The distances of the advection ($l_{\rm ad}$) and diffusion ($l_{\rm diff}$) are proportional to $B^{-1.5} \nu_{\rm m}^{-0.5}$ and $B^{-1.5}$, respectively, where $B$ is the magnetic field strength and $\nu_{\rm m} \propto E^2B$ is the photon frequency emitted by an electron of energy $E$.  By equating max($l_{\rm ad}$, $l_{\rm diff}$) to rim widths, magnetic fields were estimated to be 14--130\,$\mu$G \citep[][and references therein]{Ressler2014}.  Parizot et al.\ (2006) combined both of the advection and diffusion effects to solve for the post-shock electron distribution, finding a magnetic field of 91--110\,$\mu$G.  In 2005, Pohl et al.\ (2005) proposed the second scenario, in which several processes could lead to an exponentially decaying magnetic field downstream that limits the synchrotron distribution.  Marcowith \& Casse (2010) performed detailed calculations to test the magnetic damping model for SN~1006, finding that the data were better explained by the synchrotron-loss scenario than the magnetic damping model.  Finally, Ressler et al.\ (2014) found rapid shrinkage of rim widths with increasing photon energy that is incompatible with the magnetic damping model.  

Another observable used to infer the ambient magnetic-field strength is the precursor of either X-ray synchrotron emission and H$\alpha$ emission, whose presence is predicted by the diffusive shock acceleration theory, since accelerated electrons should spend some time ahead of the shock.  So far, no firm detection has been reported in SN~1006 \citep{Long2003,Raymond2007,Winkler2014}.  The nondetection has been considered due to amplified magnetic field upstream, since the diffusive scale length decreases with increasing magnetic field ($l_{\rm diff} \propto B^{-1.5}$ for Bohm diffusion).  By analyzing deep {\it Chandra} data of a numerous regions along the NE and SW limbs, Winkler et al.\ (2014) showed that a precursor must be thinner than $\sim$3$^{\prime\prime}$, suggesting that the magnetic field upstream is amplified to at least 45\,$\mu$G.  In addition, sophisticated nonlinear diffusive shock acceleration theories based on synchrotron losses also found the high magnetic field upstream (downstream) to be 30--40\,$\mu$G (90--130\,$\mu$G) for a large-scale ambient field of 3\,$\mu$G \citep{Berezhko2003,Morlino2010}.  

Year-scale time variabilities of synchrotron X-ray emission were found in RX~J1713.7-3946 \citep{Uchiyama2007}.  The time variabilities have been considered to be caused by a largely amplified magnetic field of a level of 1\,mG which results in extremely fast cooling/acceleration of high-energy electrons.  In contrast to these SNRs, no regions in the NE SN~1006 show strong variations in X-ray flux \citep{Katsuda2010}.  This seems to suggest low magnetic fields.  However, the absence of strong changes does not require a weak magnetic field, since steady-state particle acceleration at the shock, followed by downstream synchrotron losses, result in little flux variability.  Thus, this result is not in contradiction with the high magnetic fields inferred from the thin rims and nondetection of synchrotron precursors.  The lack of variability may be attributed to the smoothness of the synchrotron morphology, which is different from the small-scale knots in both RX~J1713.7-3946 and Cassiopeia~A.  

The strongly amplified magnetic field as well as the loss-limited thin rim interpretation support the idea that the maximum energy of high-energy electrons are limited by synchrotron cooling.  In addition, Miceli et al.\ (2013) showed that X-ray spectra from nonthermal limbs can be better explained by the loss-limited model than a simple exponential cutoff model.  Importantly, the maximum energy of electrons limited by synchrotron losses implies that the ions, which dominate cosmic rays with a ratio of ion to electron of $\sim$70 at Earth around 10\,GeV, may be accelerated to higher energies than electrons, since ions are free from synchrotron losses.  

That cosmic rays (ions) are efficiently accelerated at SNR shocks, and are not mere test particles but significantly affect shock dynamics should produce some observational signatures \citep{Reynolds2008}.  For instance, the deceleration of incoming fluid (in the shock frame), as accelerated particles diffusing ahead are scattered from incoming MHD fluctuations.  The incoming gas is gradually slowed, causing (effective) shock compression factor to increase with respect to a test-particle case.  This would result in two observational signatures: 1) a concave-up curvature in the accelerated-particle distribution, and 2) shrinkage of the gap between the forward shock and the contact discontinuity (we note, however, that Orlando et al.\ (2012) argued against this interpretation, claiming that the forward shock--contact discontinuity separation is a probe of the ejecta structure at the time of explosion rather than a probe of the efficiency of cosmic ray acceleration).  Indeed, pieces of evidence for the curved synchrotron spectrum have been accumulated by a joint spectral analysis of radio and X-ray data \citep{Allen2008} as well as broadband (radio, X-ray, and TeV gamma-rays) spectral energy distribution (SED) in a simple leptonic scenario \citep{Acero2010}.  Also, small separations between the forward shock and the contact discontinuity is seen all around the remnant, which led Cassam-Chena\"i et al.\ (2008) and Miceli et al.\ (2009) to argue efficient ion acceleration.  

\subsection{Gamma-Ray Emission}
\label{subsec:3}

A direct clue to probing high-energy ions would be to detect MeV-TeV gamma-ray emission from the decay of neutral pions resulting from inelastic collisions between high-energy ions and ambient ions (the hadronic process).  It should be noted however, that high-energy electrons can also produce gamma-rays via inverse-Compton upscattering of ambient photons (the leptonic process).  Thus, it is important to reveal contributions of both hadronic and leptonic gamma-rays, based on the intrinsic spectral difference of each emission process.  

\begin{figure}[h]
\begin{center}
\includegraphics[scale=.45]{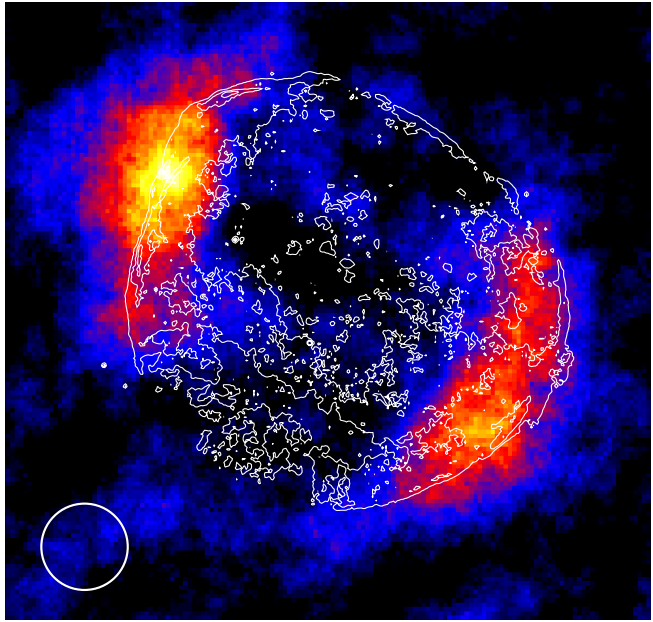}
\caption{HESS gamma-ray significance map of SN~1006 with X-ray intensity contours in white.  The white circle with a diameter of 6$^{\prime}$ in the lower left corner indicates rough spatial resolution of the HESS image.  The source of the source image was kindly provided by Fabio Acero.}
\label{fig:5}
\end{center}
\end{figure}

The High Energy Stereoscopic System (HESS) collaboration reported the first robust detection of very high-energy (TeV) gamma-rays from SN~1006 \citep{Acero2010}.  As shown in Fig.~\ref{fig:5}, the morphology of TeV gamma-ray emission strongly correlates with that of synchrotron X-ray emission.  The gamma-ray emitting region is compatible with a thin shell, although we should keep in mind that the spatial resolution of HESS (3$^{\prime}$--6$^{\prime}$) is too poor to discuss spatial scales down to a few arcminutes.  This fact suggests that the gamma-ray emission originates from the accelerated particles at the SNR shock.  

The broadband SED was, however, not very helpful to constrain the gamma-ray emission processes, so that a leptonic, hadronic, and their mixture processes could explain the SED.  From an astrophysical point of view, the magnetic field downstream must be as low as 30\,$\mu$G, in order to produce sufficient gamma-ray flux in the leptonic scenario.  Since such a field is much lower than the inference from X-ray observations (Section~\ref{subsec:2}), Berezhko et al.\ (2012) to propose that the TeV emission is dominated by the hadronic process.  The authors showed that a nonlinear kinetic model, which assumes a uniform high magnetic field of 150\,$\mu$G and comparable contributions from leptonic and hadronic processes in the TeV flux, can reasonably explain the broadband spectrum and the radial profile of TeV gamma-ray emission.  

Later, {\it Fermi} gave tight upper limits in the GeV band \citep{Araya2012,Acero2015}.  The revised SED strongly favors the leptonic scenario for the origin of TeV gamma-rays.  In this context, we may need to invoke detailed spatial variations in the magnetic field as well as the accelerated particles within SN~1006.  Petruk et al.\ (2011) constructed a leptonic model that takes into account detailed distributions of the magnetic field and high-energy particles inside the remnant, and showed that the model can explain not only the SED but also a locally high-magnetic field for the sharpest filaments in a low effective (emissivity-weighted average) magnetic field of 32\,$\mu$G.

\section{Collisionless Shock Physics}
\label{sec:5}

Shock waves are ubiquitous in a wide range of astrophysical sites including the solar wind, stellar wind bubbles, SNRs, and the merging galaxy clusters \citep[][for a recent review]{Ghavamian2013}.  In the Earth's atmosphere, the dissipation process at shocks is through direct collisions of particles, whereas in the space the density is too low to form collisional shocks, so that the dissipation is via collective interactions between particles and magnetic fields (hence, collisionless shocks).  However, dissipation mechanisms in the collisionless shocks have been still poorly understood.  For example, while one expects to find a mass-proportional temperature for each particle species, according to the Ranking-Hugoniot relation: $kT_{\rm i} = 3/16 m_{\rm i} V_{\rm s}^2$, where $m_{\rm i}$ is the particle mass and $V_{\rm s}$ is the shock speed, it has long been known that this relation does not always hold in SNR shocks.

\begin{figure}[h]
\begin{center}
\includegraphics[scale=.2]{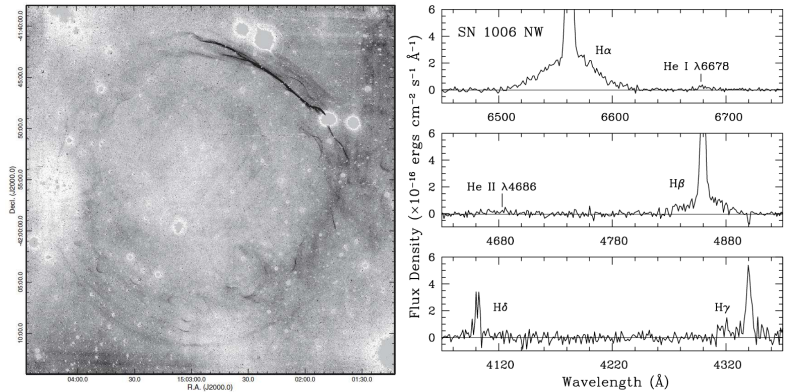}
\caption{Left: Deep H$\alpha$ image of SN~1006 after continuum subtraction, at the CTIO 4\,m Blanco telescope with the Mosaic II camera.  The NW filaments are saturated in order to reveal the faint emission elsewhere in the remnant.  This is taken from Winkler et al.\ (2014) with AAS permission.  Right: Close-up view of the optical spectrum of the NW filament.  Some prominent lines are labelled.  This is taken from from Ghavamian et al.\ (2002) with AAS permission.}
\label{fig:6}
\end{center}
\end{figure}

Measurements of post-shock temperatures have been performed based mainly on optical and UV spectroscopy of narrow filaments tracing the forward shocks in SNRs (the so-called Balmer-dominated filaments).  Figure~\ref{fig:6} left shows the entire SN~1006 taken by the 4\,m Blanco telescope at the CTIO \citep{Winkler2014}.  The brightest optical feature in this remnant is the NW filament which has been an excellent site to investigate parameters of collisionless shocks faster than 2000\,km\,s$^{-1}$.  As shown in Fig.~\ref{fig:6} right, optical emission from the shock is dominated by Balmer lines of hydrogen rather than forbidden lines of cosmically abundant elements.  Each line consists of narrow and broad components: the former is produced when cold neutrals that passed through the shock are excited in the post shock, and the latter is produced when charge-exchange reactions take place between cold neutrals and hot protons and generate a hot neutral population behind the shock.  The width of the broad component is proportional to the postshock proton temperature, while the ratio of broad to narrow flux ($I_{\rm b}/I_{\rm n}$) is sensitive to electron-proton temperature equilibration.  Ghavamian et al.\ (2002) measured the FWHM of the H$\alpha$ line and $I_{\rm b}/I_{\rm n}$ to be 2290$\pm$80\,km\,s$^{-1}$ and 0.84$^{+0.03}_{-0.01}$, respectively, which resulted in a shock speed of 2890$\pm$100\,km\,s$^{-1}$ and an electron-to-proton temperature ratio of $<$0.07.  Van Adelsberg et al.\ (2008) revised the shock velocity down to 2100$\pm$80\,km\,s$^{-1}$, by taking into account that excitation by proton collisions and charge transfer to excited levels favor the high-speed part of the neutral hydrogen velocity distribution.  A similar result was reported based on the analysis of VIMOS integral-field unit spectroscopy of the NW filament \citep{Nikolic2013}.  In this work, it was also found that broad line widths and $I_{\rm b}/I_{\rm n}$ show spatial variations across tens of atomic mean free paths, suggesting presence of suprathermal protons, the potential seed particles for generating high-energy cosmic rays.  

At UV wavelengths, Raymond et al.\ (1995) first detected line emission of He II, C IV, N V, O VI from the NW filament, using the {\it Hopkins Ultraviolet Telescope}.  Followup far-UV observations by the {\it Far Ultraviolet Spectroscopic Explore} with much better spectral resolution revealed line widths for He II, C IV, and O VI to be 2558$\pm$618\,km\,s$^{-1}$, 2641$\pm$355\,km\,s$^{-1}$, and 2100$\pm$200\,km\,s$^{-1}$, respectively.  By attributing the velocities to thermal broadening, the temperatures of He II, C IV, and O VI were found to be less than mass proportional by 48\%, 18\%, and 21\%, respectively.  This is not easy to understand at the moment, and the roles of density, pressure, magnetic field orientation, velocity, and Mach numbers should be carefully investigated to better determine the ion heating mechanisms.  

An X-ray radial profile obtained by the {\it ROSAT} High Resolution Imager peaks 8$^{\prime\prime}$ (corresponding to $\sim$0.07\,pc at a distance of 1.5\,kpc) inward from the optical filaments \citep{Winkler1997}.  Using higher spatial and spectral resolution data with {\it Chandra}, Katsuda et al.\ (2013) noticed that the offset in the radial profiles differ with energies; the 0.5--0.6 keV peak dominated by O VII is closer to the shock front than that of the 0.8--3 keV emission.  Such displacements are likely due to the fact that heavier elements need longer times to reach ionization states where they produce strong X-ray emission.  Using the reflective grating spectrometer onboard {\it XMM-Newton}, Vink et al.\ (2003) and Broersen et al.\ (2013) measured a width of O VII line in a compact knot on the NW filament, finding the velocity to be 1250$\pm$160\,km\,s$^{-1}$.  This is much slower than those of other ions measured by UV and optical wavelengths listed above, which may be explained by either Coulomb equilibration with electrons or contribution from the reverse-shocked SN ejecta or both.  

An important by-product from these shock studies is a distance, which can be derived by combining a shock velocity and a proper motion.  Using the observed values, $V_{\rm s} = 2100\pm80$\,km\,s$^{-1}$ and $\mu = 0.280\pm0.008^{\prime\prime}$\,yr$^{-1}$ \citep{Winkler2003}, the distance is estimated to be 1.57$\pm$0.07\,kpc.  This is in rough agreement with a totally independent estimate (1.45\,kpc: Section~\ref{sec:1}) based on the peak brightness of SN~1006.

\section{Conclusions}
\label{sec:6}

The present state of knowledge on SN~1006 has been reviewed.  The main conclusions are summarized below.

\begin{itemize}
\item{SN~1006 was the brightest SN witnessed in human history.  The peak brightness was likely to be around $-8$.  Several pieces of circumstantial evidence suggest that it was a Type Ia SN.  Combining the apparent magnitude with the absolute visual magnitude of normal SN Ia ($-19.12$) and the visual extinction of $A_{\rm v} = 0.31$, we can estimate the distance to be 1.45\,kpc.}

\item{No surviving companion star has been found within 4$^\prime$ of the apparent site of the explosion.  This suggests that SN~1006 originates from the DD channel rather than the SD channel.}

\item{A few background UV sources exhibit broad Fe II, Si II, Si III, Si IV absorption lines due to cold ejecta inside the remnant.  Theoretical interpretations suggest that the total mass of Fe is less than 0.16\,M$_\odot$ (assuming a spherical symmetry), which is significantly less than those for normal SNe Ia. }

\item{X-ray observations revealed that the elemental abundances in the shocked ejecta are consistent with standard theoretical nucleosynthetic models of SNe Ia.  The distribution of shocked ejecta (especially Fe) is highly asymmetric, concentrated in the SE quadrant.  The deficiency of Fe inferred from the UV absorption lines could be explained by the asymmetric SN explosion, i.e., most of Fe might have been ejected toward the SE quadrant, and escaped from the UV background sources.}

\item{SN~1006 is the first SNR in which synchrotron X-ray emission was detected from the shell of the remnant, providing us with the first evidence that SNR shocks can accelerate electrons up to $\sim$100\,TeV energies.  Thin X-ray synchrotron rims have been best interpreted as a result of rapid synchrotron cooling in a strongly amplified magnetic field downstream.  Efficient cosmic ray acceleration is inferred from pieces of observational information including the high magnetic field, a spectral curvature of synchrotron emission, and proximity between the forward shock and the contact discontinuity.}

\item{The bilateral symmetry of synchrotron emission in both radio and X-ray wavelengths (enhanced in NE and SW limbs) is most likely due to a polar cap geometry, which is strongly supported by recent radio polarization measurements as well as a strong azimuthal variation of the cutoff energy in synchrotron emission.}

\item{The morphology of TeV emission is similar to that of X-ray synchrotron emission, and the emitting region is compatible with a thin shell.  The hard gamma-ray spectrum in the GeV--TeV range suggests that the gamma-ray emission is dominated by a leptonic process at the expense of a low magnetic field of $\sim$30\,$\mu$G which conflicts to the high field of $\sim$100\,$\mu$G inferred from X-ray observations.  To self-consistently explain the thin X-ray rims and the broadband SED, we may be required to consider detailed distributions of the magnetic field and high-energy particles in the remnant.}

\item{At the NW filament that traces the forward shock, we observe extreme, but less than mass proportional, temperature nonequilibration among different species (electron, H, He, C, N, and O) behind the shock.  According to the most up-to-date shock model, the shock speed and the electron-to-proton temperature ratio can be estimated to be $\sim$2100$\pm$80\,km\,s$^{-1}$ and $<$0.07, respectively.  The shock speed, combined with a proper motion of the filament, leads to a distance of 1.57$\pm$0.07\,kpc.  This is in good agreement with that inferred from the SN brightness.}\\

\end{itemize}

\bigskip

This work is supported by Japan Society for the Promotion of Science KAKENHI Grant Numbers 25800119 and 16K17673.

\section{Cross-references}

\begin{itemize}
\item{Supernovae and supernova remnants}
\item{Historical Supernovae in the Galaxy from AD~1006}
\item{Historical records of supernovae}
\item{Supernova of 1054 and its remnant, the Crab Nebula}
\item{Supernova of 1181 and its remnant, 3C58}
\item{Supernova of 1572, Tycho's Supernova}
\item{Supernova of 1604, Kepler’s Supernova, and its Remnant}
\item{Supernova Remnant Cassiopeia A}
\item{Possible and Suggested Historical Supernovae in the Galaxy}
\item{Observational Classification of supernovae}
\item{Type Ia supernovae}
\item{Lightcurves powered by radioactivity: Type Ia and Ibc supernovae}
\item{Evolution of white dwarfs to the Thermonuclear Runaway: the smouldering phase}
\item{Dynamical Mergers}
\item{Violent Merges}
\item{Chandrasekhar mass explosions}
\item{Nucleosynthesis in thermonuclear supernovae}
\item{Dynamical Evolution and Radiative Processes of Supernova Remnants}
\item{Radio emission from Supernova Remnants}
\item{X-ray Emission Properties of supernova remnants}
\item{Ultraviolet and Optical Insights into Supernova Remnant Shocks}
\item{Infrared Emission from Supernova Remnants: Formation and Destruction of Dust}
\item{Supernova/supernova remnant connection}
\item{Supernova remnants as clues to supernova progenitors}
\item{History of using supernovae in cosmology}
\item{Peak Luminosity-Decline Relationship for Type Ia Supernovae}
\item{Low-z Type Ia calibration}
\item{Determination of H0 from supernovae}
\item{The Infrared Hubble Diagram of Type Ia Supernovae}
\item{Discovery of cosmic acceleration}
\item{Confirming cosmic acceleration}
\item{Effect of supernovae on the local interstellar medium}

\end{itemize}

%
%
%

\end{document}